\documentclass[%
 reprint,
 amsmath,amssymb,
 aps,
 longbibliography,
floatfix
]{revtex4-2}

\usepackage{graphicx}
\usepackage{dcolumn}
\usepackage{bm}

\newcommand{\xhat}{\hat{\mathbf{x}}}
\newcommand{\yhat}{\hat{\mathbf{y}}}
\newcommand{\zhat}{\hat{\mathbf{z}}}
\renewcommand{\vec}[1]{\mathbf{#1}}

\begin{document}

\preprint{APS/123-QED}

\title{Guided-Wave Sagnac Atom Interferometer with Large Area and Multiple Orbits}

\author{M. M. Beydler$^\dagger$}
\author{E. R. Moan$^\dagger$}
\author{Z. Luo}
\author{Z. Chu}
\author{C. A. Sackett}
\thanks{sackett@virginia.edu}
\affiliation{ Department of Physics, University of Virginia, Charlottesville, Virginia 22904 USA}

\date{\today}

\begin{abstract}
We describe a matter-wave Sagnac interferometer using Bose condensed atoms confined in a
time-orbiting potential trap. Compared to our previous implementation [Moan {\em et al.}, 
{\em Phys. Rev. Lett.} {\bf 124}, 120403 (2020)], our new apparatus provides
better thermal stability, improved optical access, and reduced trap anharmonicity. 
The trapping field
can be adjusted to compensate for small tilts of the apparatus in gravity.
These features
enable operation with an effective Sagnac area of 4~mm$^2$ per orbit, and we observe
interference with 25\% visibility after two orbits at a total interrogation time of
0.6 s. Long-term measurements indicate a phase stability of 0.2 rad or better.
\end{abstract}

\maketitle

\section{Introduction}
Quantum sensing is a central application in quantum science and technology. Atom interferometry is a particularly 
promising sensing technique that is useful for a wide variety of measurements, including accelerometry, gravimetry, 
and rotation sensing, as well as probing the interactions of atoms with light, surfaces, or other atoms \cite{Berman1997,Cronin2009}. 
One compelling application for atom interferometry is inertial navigation \cite{Grewal2013}, which
is important for commercial, military and space-borne vehicles that require autonomous navigation for
safety, reliability, or environmental factors. Rotation sensors are an important component of inertial navigation systems.
Compared to conventional technologies, atom interferometric rotation
sensors can provide higher sensitivity and better stability, and these benefits have been demonstrated in
laboratory experiments \cite{Gustavson2000,Durfee2006,Savoie2018}. 
However, the substantial size and complexity of these experiments limits their utility 
for practical navigation systems. One way to reduce the size of the apparatus is to use trapped atoms,
which allow for long interrogation times without the 
need for a large drop distance \cite{Wu2007,Burke2009,Moan2020}. 
While most previous interferometer implementations have used atoms in free fall,
we recently demonstrated an atomic Sagnac interferometer rotation sensor
using magnetically trapped atoms \cite{Moan2020}.
We here report improvements to our apparatus and technique, which provide a more robust system with increased reliability
and higher sensitivity. These achievements illustrate the promise of this approach for development of
a practical atom-based inertial sensor.

The Sagnac effect is applicable to any type of interferometer with arms that enclose an area $A$. If the interferometer
platform is rotating at rate $\Omega$ about an axis normal to $A$, then the net phase will include a contribution
\cite{Sagnac1913,Post1967}
\begin{equation}
\phi_S = \frac{4\pi}{\lambda v} \Omega A,
\end{equation}
where $\lambda$ is the wavelength of the interfering wave and $v$ is the phase velocity. In the case of a matter wave interferometer,
$\lambda$ is the de Broglie wavelength $2\pi\hbar/mv$ of a particle with mass $m$, leading to $\phi_S = (2m/\hbar)\Omega A$.
In comparison, an interferometer using light of frequency $\omega_L$ has $4\pi/\lambda v = 2\omega_L/c^2$, and it is the 
large ratio $mc^2/\hbar\omega_L \sim 10^{11}$ that makes atomic Sagnac interferometers attractive. However, it is necessary 
to make the enclosed area $A$ as large as possible in order to realize a high sensitivity.

Our interferometer approach was presented in Ref.~\cite{Moan2020}, and we summarize it here. A Bose condensate of $^{87}$Rb
atoms is produced in a cylindrically symmetric magnetic trap with harmonic oscillation frequency $\omega$ in the horizontal $xy$ plane.
An off-resonant standing-wave laser with wave number $k$ is applied to the condensate and, via the Bragg effect, splits it into 
two packets traveling with velocities $\pm v_B \xhat$ for $v_B = 2\hbar k/m$ \cite{Wu2005a,Hughes2007}. 
The packets move in the trap until
they come to rest at radius $R = v_B/\omega$. An orthogonal standing wave then splits the atoms into four packets with velocities
$\pm v_B \yhat$. The harmonic potential causes all four packets to move in circular orbits; the atom density is low enough
that the packets can pass through each other with negligible effect. After one or more full orbits, the $y$ Bragg beam is applied again,
which produces two interferometer outputs at $x = \pm R$. 
The Sagnac phase is differential between the two interferometers, 
while most other sources of noise are common-mode.
We observe both outputs via absorption imaging after a short time-of-flight
delay, and we extract the differential phase by plotting the two
signals against each other and fitting the points to an ellipse \cite{Collett2014}.

We note that other approaches to rotation sensing with trapped atoms are being explored. 
Atoms in a toroidal trap with tunneling junctions can exhibit rotation sensitivity analogous
to the magnetic field sensitivity of a superconducting SQUID device \cite{Eckel2014,Ryu2020}. 
Although it is difficult to achieve large Sagnac areas in this type of system, interactions between
atoms can potentially enhance the rotation sensitivity. Another approach is to implement an
atom interferometer in a linear guide that is translated perpendicular to the atomic motion
to achieve an enclosed area \cite{Wu2007}. Promising results have recently been reported using this 
method with an optical dipole trap \cite{Krzyzanowska2022}. So-called `tractor' interferometers
have been proposed in which atomic superpositions are trapped in localized spin-dependent potentials, and 
the different spin components are adiabatically moved along paths that enclose an area \cite{Navez2016,Raithel2022}.
It is also possible to implement a Sagnac interferometer using a single trapped ion \cite{Campbell2017}.
A review of compact atomic rotation
sensor techniques is available in Ref.~\cite{GarridoAlzar2019}.

In this paper, we focus on several improvements to the experiments of Ref.~\cite{Moan2020} and discuss the resulting
impact on performance. Section II details our new apparatus with improved thermal stability, optical access, 
and trap anharmonicity. Section III describes the atom interferometry methods, including the capability
to compensate for a tilted platform in gravity by adjusting the magnetic levitation force. Section IV presents
our new results, featuring multiple orbits and a total effective area of 8~mm$^2$. Finally, Section V discusses
our conclusions and directions for future development.

\section{Apparatus}

The apparatus used for Ref.~\cite{Moan2020} was detailed in Ref.~\cite{Horne2017}, 
and forms the basis for the new work here. 
It consists of a vacuum system with two chambers connected by a differential pumping tube. 
The first chamber is a cylindrical glass cell where we produce a magneto-optical trap (MOT)
of $^{87}$Rb atoms. The atoms are then loaded into a spherical dc quadrupole trap produced by an 
anti-Helmholtz coil pair. This coil pair is mechanically translated to carry the atoms into the second chamber \cite{Lewandowski2003}.
There, they are positioned at the center of a magnetic coil structure producing a time-orbiting potential (TOP) trap \cite{Petrich1995}.
We produce BEC and implement the interferometer in the TOP trap.

The original apparatus had some shortcomings that limited the interferometer performance and reliability. First,
the six trap coils were mounted on the faces of a 2-cm boron nitride cube, which was fixed to the end of a 25-cm arm
attached to a vacuum flange. The entire assembly was in vacuum, and the long arm provided a
thermal conductance of about 80 mW/K \cite{Horne2015}. The coils dissipated power of about 10~W,
leading to temperature variations of a few tens of K. These variations 
degraded the stability of the trap, resulting in phase noise for the interferometer \cite{Luo2021}
and requiring experimental adjustments on the one-hour time scale.
Second, optical access for one of the Bragg lasers was achieved by passing
the beam along the axis of the 2-m long vacuum system, including through a 9-mm diameter hole in the coil-mounting arm
and the 12-mm diameter tube connecting the two chambers. This severely restricted the adjustability
of the beam alignment. Third, the coil geometry introduced tilts and anharmonicity to the trap
potential, which resulted in the wave packets failing to overlap after more than one orbit.

\begin{figure}
\includegraphics[width=\columnwidth]{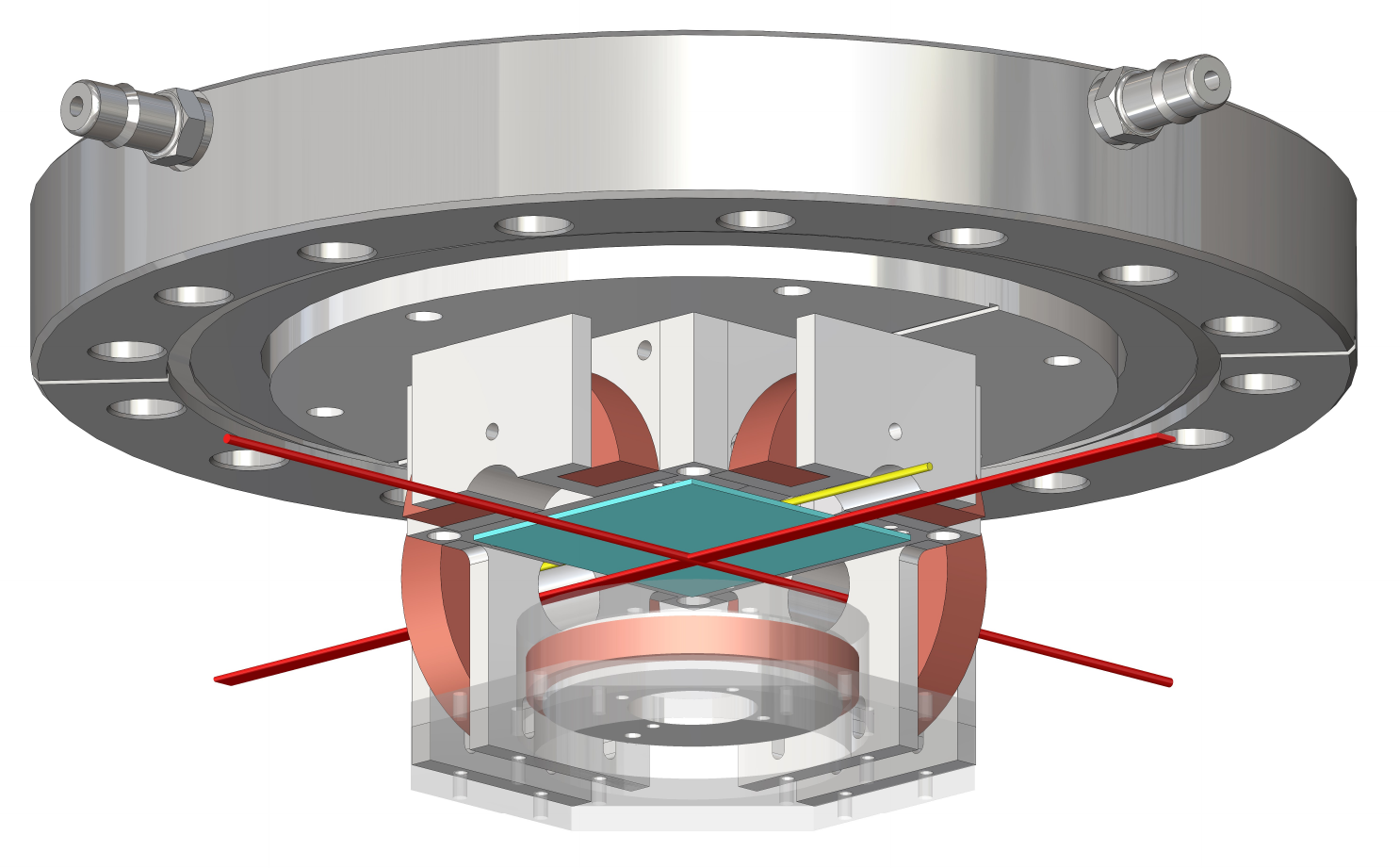}
\caption{Cutaway view of coil fixture and cooling flange. The gray component at the top of the image is a six-inch 
conflat flange with water-cooling connectors. The white components are Shapal forms for the
coils, which are represented in copper color. There are six coils altogether, of which five are visible.
The sixth is enclosed by the rectangular Shapal base above the blue plate, which is 
a 1-mm thick silicon wafer used as a mirror for imaging.
The four horizontal coils have 64 turns of AWG 20 copper wire with Kapton insulation. The average
radius is 14 mm and the average distance from the fixture center is 26 mm. The two vertical coils have
26 turns of AWG 20 wire, an average radius of 15 mm, and average distance of 14 mm. The yellow wire
is an antenna for rf evaporative cooling. The red beams show the paths of the two Bragg
lasers. Atoms are trapped at the center of the fixture, about 2 mm from the mirror surface.
\label{fixture}}
\end{figure}

The new apparatus described here addresses these shortcomings. To provide better thermal management,
the trap coils are now larger and are mounted in a fixture that is directly attached to
a water-cooled flange, as illustrated in  Fig.~\ref{fixture}. The coil fixture is electrically insulating
so that the TOP fields do not induce eddy currents. It is machined from Shapal Hi-M Soft,
an aluminum nitride ceramic with a thermal conductivity
of 90~W/(m$\cdot$K). The six coils are wrapped on Shapal forms that are bolted and epoxied to a Shapal base.
The base is then screwed to a stainless-steel CF6 conflat flange, which was machined to a thickness of 8.5~mm. 
A 3-mm deep serpentine pattern is machined on the air side of the flange for water cooling, and the 
pattern is covered by a thin steel cap bolted to the flange and sealed with silicone adhesive. The new system has a measured
thermal conductance of 2~W/K. The power dissipated in the coils
remains about 10~W, so the total temperature change is reduced to about 5~K.
However, we minimize the temperature variations by running the experiment at a steady duty cycle,
and we estimate the resulting variation to be about 1~K from run to run.

Thermal fluctuations can affect the trap in two ways. 
First, thermal expansion causes the coil
geometry to change, which alters the magnetic fields and therefore the trapping potential. 
Since the magnetic field from a source varies inversely with distance, we have
$|(dB/dT)/B| \approx |(d\ell/dT)/\ell|$, where $B$ is the field magnitude, $T$ is the temperature, 
and $\ell$ is the fixture size scale. The relative size variation $(d\ell/dT)/\ell$ is determined
by the coefficients of thermal expansion for copper and Shapal, which are $2\times 10^{-5}$ K$^{-1}$ and
$5\times 10^{-6}$ K$^{-1}$, respectively. The second effect is the change with temperature of the resistance 
of the copper wires, which alters the
driving current. Our trap drive electronics use a current-stabilization feedback loop
to counter such effects, but the loop gain of about 30~dB provides only a factor of
33 reduction \cite{Baranowski2006}. The temperature coefficient for copper is $4\times 10^{-3}$~K$^{-1}$, so 
the expected relative current variations are about $10^{-4}$~K$^{-1}$. The relative field
variation is the same, so we expect this resistance effect to dominate over the thermal expansion effect. Given the estimated
temperature variations, we expect the trap potential to be stable to approximately one part
in $10^4$. This is similar to the amplitude stability of the function generators from
which the current signals are sourced
\cite{Baranowski2006}. These improvements reduce phase noise in the rotation measurement and
allow the experiment to run with minimal intervention for multiple days.

In the new system, the two Bragg laser beams enter the vacuum chamber through CF133 conflat viewports,
providing improved optical access.
The viewports are anti-reflection coated to give 0.25\% reflection per surface. However, the experiment
is sensitive even to these suppressed reflections since they can cause significant spatial
variation in the laser intensity. To suppress this effect, the copper gaskets used to
mount the viewports are machined with a 3 deg wedge angle, so that the windows are 
off-normal to the laser. We did not find the wedge 
to have any effect on the vacuum seal.

The new coil arrangement imposes changes to our imaging system. We monitor the
output of the interferometer using a vertical absorption imaging beam.
Because of the water-cooling flange, it is no longer possible to pass this beam through
the chamber. Instead, we mount a flat mirror on the coil fixture above the atoms, 
from which the absorption laser reflects. 
The absorption imaging system produces a picture with two atom clouds: one in focus 
and the other one out of focus. This ``double image'' is due to the roughly 2 mm 
distance between the mirror surface and the BEC. 
We find that the image analysis works best if the absorption beam is
tilted off of vertical by enough to displace the in-focus and out-of-focus absorption
features in the image, as can be seen in Fig.~\ref{AIimage}. 
For the data discussed here, 
the in-focus feature is fit to a Gaussian profile, and the out-of-focus feature is 
ignored. Potentially, the imaging performance could be improved by fitting to both
features. 

A second imaging system uses a horizontal
absorption beam to observe the atom clouds' vertical positions. The horizontal imaging
can use either Bragg beam axis, with the imaging light directed to the camera
via a beam splitter. In the previous apparatus, polarizing optics were used to separate
the Bragg and imaging beams, but we found this led to increased intensity fluctuations
due to imperfect polarization coupling in the optical fibers that deliver the beams.

The process for producing Bose condensates is similar to that of the former
apparatus. About $10^9$ $^{87}$Rb atoms are collected in the MOT,
optically pumped into the $F=2$, $m=2$ Zeeman state, loaded into a dc spherical quadrupole
magnetic trap with $dB/dz \approx 350$~G/cm, 
and then transported to the new chamber using coils mounted on a 
motorized translation stage \cite{Lewandowski2003}.
Atoms enter the coil fixture through a gap between adjacent horizontal coils. 
An initial stage of rf-induced evaporative cooling is performed
in the dc quadrupole trap, cooling the atoms from a temperature of about 1~mK to about
100~$\mu$K, where Majorana losses start to become significant. Oscillating currents
are then applied to the fixture coils to produce a TOP trap. The initial bias field
amplitude is about 40~G. Over 9 s, the amplitude is reduced to 5~G, which compresses
the trap and evaporatively cools the atoms via Majorana losses at the radius of the orbiting zero.
After compression, the trap confinement frequencies are about 100~Hz. 
A second stage of rf evaporation is then applied, leading to nearly pure
Bose condensates with typically $2\times 10^4$ atoms. 

The atoms are adiabatically transferred from this relatively tight trap to a weakly confining
trap suitable for atom interferometry. The nominal field in the final trap is \cite{Moan2020}
\begin{align} \label{TOPfield}
\vec{B} & = B_0\left[\sin\Omega_1 t\cos\Omega_2 t \xhat
    +\sin\Omega_1 t \sin\Omega_2 t \yhat
    +\cos\Omega_1 t \zhat\right] \nonumber \\
 &    +\frac{1}{2}B_1' \cos\Omega_1 t (x\xhat + y\yhat -2z \zhat),
\end{align}
where $B_0$ is the bias amplitude and $B_1'$ is the gradient amplitude. The two TOP frequencies are
$\Omega_1 = 2\pi\times 10$~kHz and $\Omega_2 = 2\pi\times 1$~kHz. The resulting time-averaged
potential energy $\mu\langle |B|\rangle$ provides harmonic confinement along with a
a linear term $-\mu B_1'z/2$ that we set equal to $mgz$ to cancel gravity. Here $\mu \approx \mu_B$ is
the atomic magnetic moment and $g$ the acceleration of gravity.
The weak trap is established by first ramping up the $B_1'$ ac gradient amplitude over
1~s, and then ramping down the dc quadrupole coil current over the course of 18 s. As the dc quadrupole
is ramped off, the bias amplitude $B_0$ is also gradually increased from 5~G to 15~G.
By carefully adjusting the dc trap location and the parameters of the ramps, 
the residual oscillation amplitude of the atoms in the final trap is limited to about
30~$\mu$m. For the work presented here, the trap has horizontal frequencies near 3.5 Hz and 
a vertical frequency near 4.0 Hz. In comparison, the work of \cite{Moan2020} used trap frequencies
near 10~Hz.

While the coil fixture is electrically insulating, the TOP fields do
produce eddy currents in the steel vacuum chamber and mounting flange. These
alter the applied fields by roughly 10\%. We have not attempted to model the eddy currents,
but instead compensate for their effects experimentally by adjusting the coil driver
amplitudes and phases. 

\begin{figure*}[t]
\includegraphics{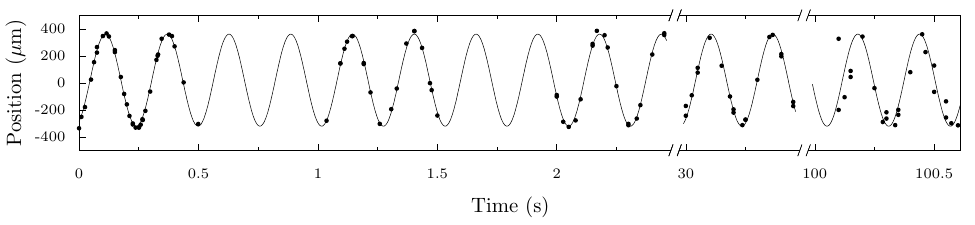}
\caption{Position of atoms in TOP trap. Data points show the central position of
a single condensate as it oscillates vertically in the trap; note the breaks in the horizontal scale.
The curve is a single sinusoidal fit to the full data set, including points not shown at intermediate times.
It can be seen that the atomic oscillations remain
coherent up to a time of 100~s. The fitted frequency is $\omega_z = 2\pi\times 3.8673(3)$~Hz.
  \label{longosc}}
\end{figure*}

To characterize the stability of the trapping potential, we deliberately
excited a collective `sloshing' oscillation of the condensate
and measured the time scale over which the oscillations remained coherent. Representative
data are shown in Fig.~\ref{longosc}. The oscillations
developed about 0.2 rad of phase noise after 100~s. This corresponds to a relative stability of
$10^{-4}$, in agreement with expectations based on our thermal analysis.

Oscillation data is also useful for characterizing the anharmonicity 
of the trap. Anharmonicity is important because it can cause both uncontrolled phase
shifts and a loss of visibility for the atom interferometer \cite{West2019,Luo2021}.
The most significant
anharmonic contributions in our case give a potential
\begin{align} \label{eq:anharm}
V(\rho,z) = &  \frac{1}{2}m\omega^2\bigg(\rho^2 + \zeta^2 z^2 \\ \nonumber
& +a_{21}\frac{\rho^2z}{R} + a_{03}\frac{z^3}{R}  + a_{40}\frac{\rho^4}{R^2}  + a_{04}\frac{z^4}{R^2} \bigg),
\end{align}
where $\rho^2 = x^2 + y^2$ and $\zeta = \omega_z/\omega$. We characterize the anharmonic terms  
by dimensionless parameters $a_{\mu\nu}$, with the orbit
radius $R$ serving as a scale factor. This facilitates comparison between traps with different confinement strengths.

We determine the values of the $a_{\mu\nu}$ by analyzing trajectories of atoms oscillating
in the trap as in Ref.~\cite{Moan2020a}. We use a classical approximation in which a wave packet is modeled as a
Newtonian particle moving in the anharmonic potential. Table 1 compares the
results of this analysis for our old trap with $R = 0.20$~mm and our current trap at $R = 0.53$~mm. 
We find that even with the increased radius, the new trap is as good as or better than the previous version.

\begin{table}
\setlength\tabcolsep{1.5ex}
\renewcommand{\arraystretch}{1.5}
\begin{tabular}{l|cccc}
  & $a_{21}$ & $a_{03}$ & $a_{40}$  &  $a_{04}$  \\ \hline
 Old trap & 0.10(30) & 0.090(10) & -0.0030(3) & 0.016(5)  \\
 New trap & 0.12(6) & 0.047(5) & -0.0036(3) & 0.006(3)  
 \end{tabular}
 \caption{Anharmonic parameters of the old and new TOP traps, as defined in Eq.~\protect\eqref{eq:anharm}.
  The value of $a_{04}$ in the old trap is corrected for an error in Eq.~(21) of Ref.~\protect\cite{Moan2020a}.} 
 \end{table}


\section{Methods}


The condensate wave packets in the new trap have Thomas-Fermi size $L \approx 20~\mu$m,
while the Sagnac orbit radius is $R \approx 500~\mu$m. The dual interferometer
requires that both sets of packets overlap with each other after one or more orbits,
and this requires precise control of both the Bragg beam alignment and the trap
potential: the two Bragg lasers must be aligned to be horizontal and orthogonal to each other, 
the trap must be adjusted to be cylindrically symmetric, and the gradient
of the trap must be aligned parallel to gravity. We outline here the procedures
used to achieve these conditions.

The Bragg beams operate at a typical detuning of 20~GHz from the D2
resonance at 780.24~nm. The $y$ Bragg beam is shaped by anamorphic prisms to 
give Gaussian waists of 0.6~mm$~\times$~2.5~mm, which improves its uniformity
across the two interferometers. The $x$ beam is applied only to a single packet, so 
it is circular with a waist of 1.1~mm. The two beams are aligned
to be orthogonal to an accuracy of about 1 deg by splitting stationary condensates 
and observing the resulting trajectories with the vertical imaging camera. 
The vertical beam alignment is adjusted by observing the trajectories with
the horizontal camera. The $\rho^2 z$ anharmonic term is large enough to make
the horizontal motion deviate significantly from a straight line, but the Bragg
beam angle can be set with an accuracy of 1 deg by ensuring that the trajectories
of the two split packets are initially horizontal.

The cylindrical symmetry of the trap depends on both harmonic and anharmonic
terms. In this work, we did not attempt to control anharmonic asymmetries. 
Harmonic asymmetries in the potential are parameterized by dimensionless
$\Delta$ and $\gamma$ as 
\begin{equation} \label{harmonic}
V = \frac{1}{2}m\omega^2\left[(1+\Delta) x^2 + (1-\Delta) y^2 + 2\gamma xy\right].
\end{equation}
Trap asymmetries can arise from imperfect coil windings, fields from lead wires, and induced eddy currents.
As discussed in \cite{Moan2020}, our TOP trap configuration allows these terms to be corrected
using the current drive signals. The diagonal asymmetry $\Delta$ can be controlled by adjusting
the relative amplitudes of the rotating bias field components along $x$ and $y$,
with $\Delta = (B_x-B_y)/7B_0$ and mean bias amplitude $B_0 = (B_x + B_y)/2$. 
The off-diagonal
term $\gamma$ is controlled by the phase offset $\beta$ between the
$x$ and $y$ fields, with $\gamma = 2\beta/7$ \cite{Luo2021}. The amplitude and phase parameters
are initially adjusted by observing the oscillating trajectories of split condensates
in the $x$ and $y$ directions. A non-zero $\Delta$ value causes the oscillation
frequencies to be unequal, while a non-zero $\gamma$ value causes the
trajectories to curve away from the initial axis. 

In practice, the $\gamma$ parameter and the vertical beam alignment are 
the most important of these imperfections for the interferometer
performance. The off-diagonal asymmetry causes the final positions of the wave packets to separate 
horizontally by
a distance $\Delta x = \pi n\gamma R$ after $n$ orbits. 
Maintaining good overlap $\Delta x \ll L$ thus requires $n\gamma \ll L/\pi R \approx 10^{-2}$.
The vertical angle $\xi$ of the $y$ Bragg beam causes the two packets to be displaced
vertically by an amount $\Delta z = (2 R \xi/\zeta ) \sin (2\pi n\zeta)$,
so ensuring good overlap requires $\xi \ll L\zeta/2R \approx 2\times 10^{-3}$ rad. The trajectory measurements
are typically sufficient to reach $\Delta y, \Delta z \approx L$ so that interference can be observed,
and then further fine adjustments are made to maximize the interference visibility.

\begin{figure}
\includegraphics[width=3.5in]{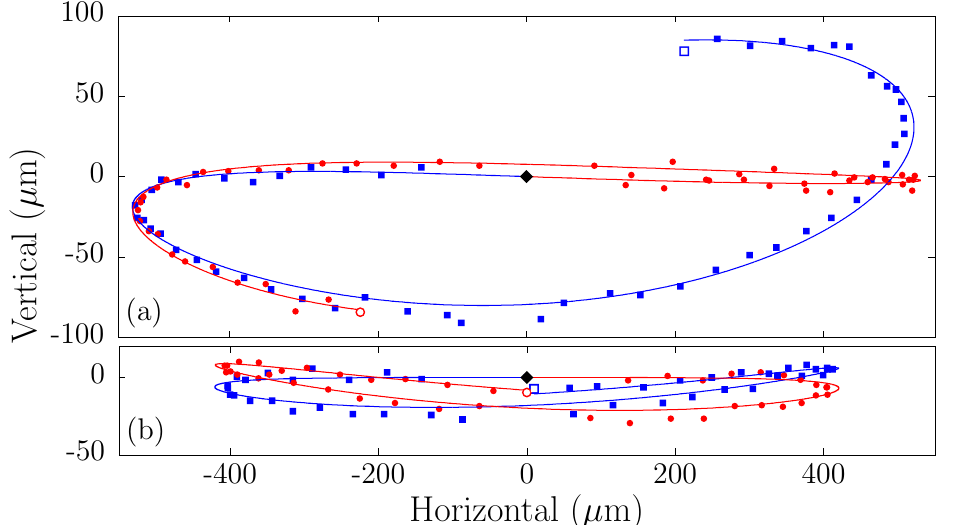}
\caption{Atomic trajectories in the presence of anharmonicity. In each graph, the red circles and blue squares show the trajectories
of wave packets launched to the right and left, respectively. The black diamond indicates the common starting point at the trap center,
and the open symbols indicate the final point of each trajectory. Points give the center location of the wave packet, determined by
analyzing absorption images at different times of flight. Curves are model trajectories for
a Newtonian particle in the anharmonic potential. (a) Asymmetric trajectories observed when the vertical trap axis is tilted by approximately 12 deg due to eddy currents in the vacuum chamber walls. Here the two wave packets fail to overlap after
a complete oscillation.
(b) Symmetric trajectories obtained by modifying the field as in Eq.~\protect\eqref{tilteq} with $\epsilon \approx 0.7$, $\psi \approx 50$ deg. Here the packets are well overlapped after a complete oscillation. 
The relatively large value of $\epsilon$ used also causes an increase in the horizontal oscillation
frequency, from 3.3~Hz to 4.3~Hz.
\label{tiltfig}}
\end{figure}

A final adjustment that we found to be critical is the alignment of the trap potential gradient with gravity. The field of Eq.~\eqref{TOPfield}
provides a vertical gradient and horizontal symmetry, but we initially observed
asymmetric trajectories such as in Fig.~\ref{tiltfig}(a), where the vertical motion is different
for packets receiving positive and negative horizontal Bragg kicks. Modeling of the trajectories
indicated that the magnetic gradient was in fact a few degrees off of vertical, which we attribute
to the effects of eddy currents in the vacuum chamber. In a perfectly harmonic trap, 
a misaligned gradient would simply displace the trap center, but in the presence of the relatively
large $\rho^2 z$ anharmonicity, it introduces asymmetry that prevents the interferometer trajectories from closing.

In order to correct for this effect, we 
modify the gradient time dependence in \eqref{TOPfield} to 
\begin{equation} \label{tilteq}
	\frac{1}{2}B_1'\big[\cos\Omega_1 t + \epsilon \sin\Omega_1 t \cos(\Omega_2 t + \psi)\big](x\xhat + y\yhat -2z \zhat),
\end{equation}
where $\epsilon$ and $\psi$ are variable control parameters. Because the gradient now has a component
matching the time-dependence of the horizontal bias fields, the time-averaged field 
acquires a horizontal gradient that evaluates to first order in $\epsilon$ as
\begin{equation}
	\frac{1}{8}\epsilon B_1'(x\cos\psi -y\sin\psi).
\end{equation}
We then adjust $\epsilon$ and $\psi$ to symmetrize the wave packet trajectories, as
seen in Fig.~\ref{tiltfig}(b). While we use this technique here to correct for the effects
of eddy currents, it could more generally be applied to compensate for the motion of a
vehicle in an inertial navigation system.

\section{Results}

\begin{figure}
\includegraphics[width=0.75\columnwidth]{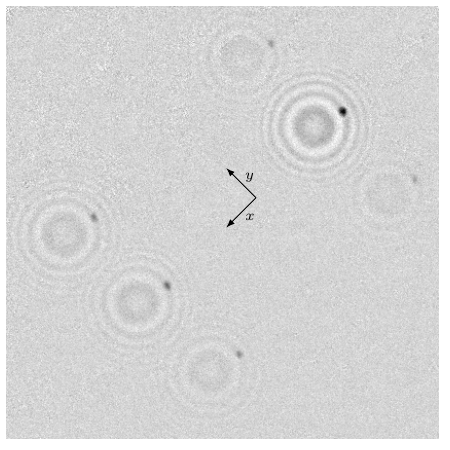}
\caption{Example interferometer output image. The three absorption features in the lower left are the atoms
from interferometer 1 with a signal $S_1 = 0.74$, and the three features in the upper right are from 
interferometer 2 with signal $S_2 = 0.43$. Each atomic wave packet produces two images, one in focus
and seen as a compact spot, and the other out of focus and seen as a set of rings. These arise because 
the absorption probe beam passes through the atoms twice; see Fig.~\protect\ref{fixture}.
The coordinate axes
indicate the trap center (relative to the compact spots)
and the Bragg beam directions. The field of view shown is 1.56~mm across. 
\label{AIimage}}
\end{figure}

\begin{figure}
\includegraphics[width=0.85\columnwidth]{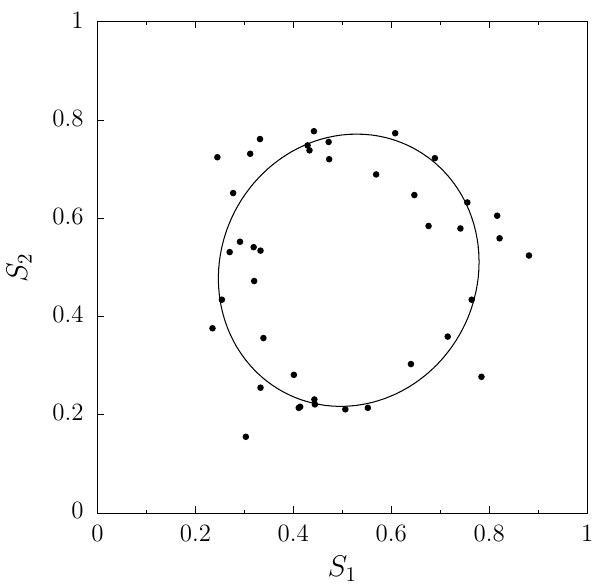}
\caption{Interference ellipse, single orbit. Data points are signal $(S_1, S_2)$ pairs obtained in $N = 40$ sequential
runs of the experiment. The curve is a fitted ellipse yielding a differential phase $\Phi = 1.5(1)$~rad
and visibilities $V_1 = 0.52$, $V_2 = 0.56$.
\label{ellipsefig}}
\end{figure}

Figure~\ref{AIimage} shows a typical absorption image of the interferometer output, consisting of three output wave packets for each
of the two interferometers. Each of the six absorption peaks is fit to a Gaussian to determine the relative atom numbers
$N_{ij}$, where $i = 1$ or 2 labels the two interferometers, and $j = +, 0$ or $-$ labels the output momentum. For each
interferometer, an output signal is calculated as
\begin{equation}
S_i = \frac{N_{i0}}{N_{i0} + N_{i+} + N_{i-}}.
\end{equation}
The signals depend on the interferometer phases as $S_i = S_{i0}(1 + V_i \cos \phi_i)$, where $S_{i0} \approx 1/2$ is the
mean signal, $V_i$ is the interferometer visibility, and $\phi_i$ is the phase. The individual phases are randomly 
distributed between $0$ and $2\pi$, primarily due to vibrations of the optical table.
The fluctuations are correlated, however, with $\phi_2 - \phi_1 \equiv \Phi$ approximately constant. To extract
$\Phi$ we plot $S_2$ vs.\ $S_1$ and fit the resulting points to an ellipse, as seen in Fig.~\ref{ellipsefig}. 
The differential phase is expressible as
\begin{equation}
\tan \Phi = \frac{2ab}{(a^2-b^2)\sin2\theta},
\end{equation}
where $a$ is the ellipse major axis, $b$ is the minor axis, and $\theta$ is the angle of the major axis. Here
$\Phi$ and $2\pi-\Phi$ yield equivalent ellipses, so we obtain a phase in the range $0\leq\Phi \leq \pi$.  

For the data set in Fig.~\ref{ellipsefig}, we obtain $\Phi = 1.51$~rad. The expected Sagnac phase
from the Earth's rotation is 0.5 rad, but trap imperfections and anharmonicities also contribute to $\Phi$.
Using the trap parameters and uncertainties reported here with the semiclassical phase calculations of Ref.~\cite{Luo2021},
we estimate the total trap phase to fall in a range $\pm 20$~rad.  The static value we obtain for $\Phi$ is 
therefore not directly useful for rotation sensing, but to the extent that the trap phase is stable, the
interferometer can measure changes in rotation rate. This is not a limitation for navigation applications
since the zero-rotation bias can be calibrated and subtracted out.

The sensitivity to rotations depends on the uncertainty in the phase.
To estimate the uncertainty $\sigma_\Phi$, we numerically calculate the variation in $\Phi$
required to increase the $\chi^2$ goodness-of-fit parameter by a factor of $1/(N-5)$, where
$N$ is the number of data points and 5 is the number of fit parameters \cite{Press1992}. As $\Phi$ is varied,
the other four fit parameters are adjusted to keep $\chi^2$ minimized. For the data set of Fig.~\ref{ellipsefig},
we find $\sigma_\Phi = 0.1$~rad. In comparison, when we acquire multiple data sets under the same
conditions, we obtain a standard deviation of 0.15~rad, in reasonable agreement.

The standard quantum limit (SQL) for the differential phase uncertainty is $2/N_\mathrm{tot}^{1/2}$, where
$N_\mathrm{tot}$ is the total number of atoms measured over the 40 runs. Here $N_\mathrm{tot} \approx 10^6$,
so the SQL is 2~mrad, about 50 times lower than observed. Part of this discrepancy comes from the ellipse fitting
process: because the phase of each individual interferometer is not controlled, many measurements
occur near $\phi_i = 0$ or $\pi$ where the phase sensitivity is low, so even a perfect 
interferometer of this type would not achieve the standard quantum limit. However, our analysis
suggests that shot-to-shot technical phase noise of order 1~rad is also present, which we attribute to
variations in the trap phase caused by fluctuations in the TOP fields \cite{Luo2021}.
An investigation of the noise performance of the interferometer is ongoing and will be the subject of a 
future paper.

We attribute the imperfect visibility of the interferometers to tilting of the trap, which is challenging to 
set precisely. Although the field-tilting method of Eq.~\eqref{tilteq} is precisely adjustable, 
changing the trap tilt causes the atoms to move relative to the coil structure
and Bragg beams, which then requires considerable readjustment of other parameters. We estimate that we achieve a tilt
precision of about 1 deg, which corresponds to a packet overlap accuracy of about 20~$\mu$m. 
Since this is comparable to the packet size, a visibility of 0.5 is plausible.

\begin{figure}
	\includegraphics[width=0.85\columnwidth]{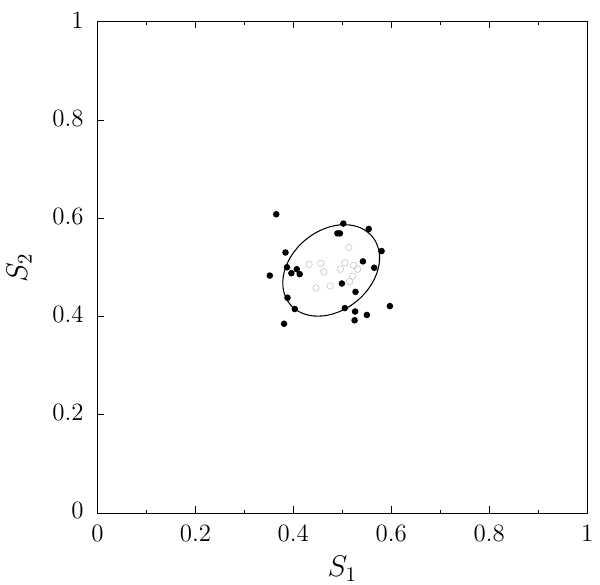}
	\caption{Interference ellipse, two orbits. The solid black points are $(S_1, S_2)$ values as in Fig.~\protect\ref{ellipsefig},
		and the curve is an elliptical fit yielding $\Phi = 1.3(2)$, $V_1 = V_2 = 0.19$. The open gray points illustrate detection noise
		and are acquired by deliberately misaligning the the interferometer to suppress interference. The open points have been
		uniformly offset to the center of the ellipse curve, to improve clarity. 
		\label{twoorbitfig}}
\end{figure}

With the single-orbit interferometer optimized, we were able to observe an interference signal
after two orbits as well, with results shown in Fig.~\ref{twoorbitfig} (solid black points). The visibilities are reduced to
0.19 for each interferometer, but the ellipse shape of the graph is still clearly distinguishable.
For operation with three orbits, we were unable to obtain interferometer data
that was distinguishable from noise.

The open gray points in Fig.~\ref{twoorbitfig} are measurements
obtained when the trajectories are deliberately misaligned so that the wave packets fail to overlap,
illustrating detection noise in the system. The standard deviation of the signals is then
$\sigma = 0.03$, which can be compared to the standard quantum limit of 0.01 for a single measurement. 
We attribute the excess noise to
spatial variations in the absorption probe and reference laser beams.

The data of Figs.~\ref{ellipsefig} and \ref{twoorbitfig} were taken in a trap with horizontal frequency $\omega = 2\pi\times 3.3$~Hz 
and orbit radius $R = 0.57$~mm. The enclosed area for one orbit is thus 1.0~mm$^2$, but the effective area accounting for
all four packets is 4.1~mm$^2$ for the single-orbit measurement and 8.2~mm$^2$ for the two-orbit results. Although we did not
apply a deliberate rotation in these experiments, we calculate the Sagnac sensitivity
\begin{equation}
\frac{\Phi_S}{\Omega} = \frac{8\pi n m R^2}{\hbar},
\end{equation} 
where $n$ is the number of orbits.
The experiment of Fig.~\ref{ellipsefig} has a sensitivity of 
$1.1\times 10^4$ rad/(rad/s) and a phase uncertainty $\sigma_\Phi = 0.1$~rad, yielding
a rotation error of 9~$\mu$rad/s. The two-orbit data of Fig.~\ref{twoorbitfig} has 
twice the sensitivity but also twice the uncertainty, leading to the same rotation error.
In either case, this is an order of magnitude improvement
over the results reported in Ref.~\cite{Moan2020}.

We also characterize the long-term stability of the single-orbit interferometer. Over a period of 24 hours, we observe variations
$\delta\Phi$ of approximately 0.2 rad, which are consistent with the short-term phase noise. Similar variations were 
observed either with the interferometer running continuously for the 24-hour period, or with the experiment shut down
and restarted 24 hours later. In the second test, the lasers, trap currents, and Rb source were turned off, while 
vacuum pumps and other control equipment were left on. After restarting, the apparatus was allowed to warm up for 
1 hour before interferometer measurements were taken. In both tests, the only user interventions applied were 
adjustments to the Bragg laser power to maintain optimum splitting pulse efficiency, with adjustments of a few
percent required every few hours. The measured phase variations correspond to a rotational bias stability
better than 20~$\mu$rad/s. Further investigations of bias stability will also be reported in future work.

\section{Conclusions and Future Directions}

Compared to our previous system, the apparatus presented here 
demonstrates roughly an order of magnitude improvement in both trap stability and
rotation sensitivity. The system is robust and reliable, opening up new 
techniques like multiple orbits and tilt control, and allowing for investigations of noise and 
drift.

However, further improvements will be required to be potentially useful for inertial 
navigation. The sensitivity of a navigational rotation sensor
can be expressed using the angle-random walk (ARW) parameter $\delta\Omega\, \tau^{1/2}$, where
rotational accuracy $\delta\Omega$ is obtained in averaging time $\tau$ \cite{Grewal2013}. A typical
MEMS gyroscope has an ARW on the order of $5$ deg/h$^{1/2}$, while a high-performance
optical gyroscope can achieve $10^{-3}$ deg/h$^{1/2}$ \cite{Woodman2007}. Given the 4000 s measurement time
required to take the data of Fig.~\ref{ellipsefig}, our system exhibits an ARW of 
$2$ deg/h$^{1/2}$. Potential improvements that can be considered 
include eliminating technical phase noise,
decreasing the cycle time of the experiment from 100~s to 5~s using atom-chip or 
optical-trap techniques \cite{Farkas2014}, increasing the single-orbit area to 1.5~mm$^2$, 
and increasing the number of orbits to 10. All together, these
would result in an ARW parameter of $10^{-4}$~deg/h$^{1/2}$, which would be compelling
for a compact sensor. Although these improvements are technically challenging,
they are compatible with fundamental constraints including collisional losses and 
dephasing effects from interactions. 

The bias stability in navigation applications is also important \cite{Grewal2013}. It 
ranges from tens of degrees per hour for MEMS devices to $10^{-3}$ deg per hour for optical
sensors \cite{Woodman2007}. Our observations indicate stability better than 4~deg/h, but this is likely
limited by short-term noise rather than long-term drift. Good 
stability is typically a feature of atomic devices, since atoms are intrinsically
stable and highly effective laser stabilization techniques are available. 
However, our interferometer is also sensitive to the trapping fields, which
are subject to drifts. For instance, one candidate for the short-term phase
noise that we observe is fluctuations $\delta\omega$ in the
trap frequency coupled to a non-zero $\gamma xy$ term in the trap potential of
Eq.~\eqref{harmonic}. From Ref.~\cite{Luo2021}, this leads to phase fluctuations
\begin{equation}
\delta\Phi = 2\pi^2 kR\gamma (n+2n^2) \frac{\delta\omega}{\omega}.
\end{equation}
If $\gamma$ can be stabilized at the $10^{-4}$ level and long-term
variations $\delta\omega/\omega$ controlled at the $10^{-6}$ level,
then the expected phase noise would be $3\times 10^{-5}$~rad for the 
single-orbit parameters demonstrated here, and $5\times 10^{-3}$~rad in the 
high-sensitivity system outlined above. 
The corresponding rotational bias stability would be $5\times 10^{-4}~$deg/h and
{$1\times 10^{-4}$~deg/h, respectively. These performance levels again seem feasible,
if challenging.

Our primary motivation for pursuing a trapped-atom interferometer is
the potential for a compact apparatus, as size requirements are
in practice critical for inertial navigation applications. The vacuum chamber
in our present apparatus is about 2 m long and is installed on a conventional 
optical table. However, Bose condensation can be achieved in much smaller 
devices using atom-chip or laser-trap technologies. Compact condensate interferometers
have been implemented both on sounding rockets \cite{Muentinga2013} and on the International 
Space Station \cite{Aveline2020}, with total
system volumes of about $10^6$~cm$^3$. Advances in integrated photonics \cite{Pelucchi2021}
and vacuum technology \cite{Burrow2021}
are expected to reduce these requirements significantly.
In comparison, a high-performance conventional inertial navigation system requires volume of order
$10^4$~cm$^3$ \cite{Kayton1997}. 
While engineering advances will be required to make the atomic systems
competitive, the improvements do not seem out of reach.

In total, we hope the advances reported here illustrate that the trapped atom interferometer 
technique has realistic potential for applications in inertial navigation,
and that further efforts in this area are warranted.

\acknowledgments
We acknowledge technical contributions from
Seth Berl, Evangeline Sackett, and Malcolm Schlossberg,
and productive discussions with Eric Imhof, Evan Salim,
and James Stickney. We thank Elizabeth Larson for helpful comments on the manuscript.
This work was supported by the National Science
Foundation (Grant No. PHY-2110471), NASA (Contract
No. RSA1662522), and DARPA (Grant No. FA9453-19-1-0007). 

$^\dagger$M.B. and E.R.M. contributed equally to this work.

\vspace{2ex}

\noindent {\bf AUTHOR DECLARATIONS}

\vspace{2ex}

\noindent {\bf Conflict of Interest}

The authors have no conflicts to disclose.

\vspace{2ex}

\noindent {\bf DATA AVAILABILITY}

The data that support the findings of this study are available from the corresponding author
upon reasonable request.

%

\end{document}